\documentclass{article}
\usepackage{spconf,amsmath,graphicx}
\usepackage{amsthm,amsmath}
\usepackage{amsfonts,amssymb}
\usepackage{subfigure}

\title{Foster Strengths and Circumvent Weaknesses: a Speech Enhancement Framework with Two-branch Collaborative Learning
}
\name{Wenxin Tai*, Jiajia Li*, Yixiang Wang, Tian Lan$\dagger$\thanks{* denotes authors have same contributions. This research has been supported in part by the National Science Foundation of China (U19B2028, 61772117), National Science and Technology Commission Innovation Project (19-163-21-TS-001-042-01),  and the Fundamental Research Funds for the Central Universities (ZYGX2019J077).}, Qiao Liu}
\address{University of Electronic Science and Technology of China}
\begin{document}
%
\maketitle
\begin{abstract}
Recent single-channel speech enhancement methods usually convert waveform to the time-frequency domain and use magnitude/complex spectrum as the optimizing target. However, both magnitude-spectrum-based methods and complex-spectrum-based methods have their respective pros and cons. In this paper, we propose a unified two-branch framework to foster strengths and circumvent weaknesses of different paradigms. The proposed framework could take full advantage of the apparent spectral regularity in magnitude spectrogram and break the bottleneck that magnitude-based methods have suffered. Within each branch, we use collaborative expert block and its variants as substitutes for regular convolution layers. Experiments on TIMIT benchmark demonstrate that our method is superior to existing state-of-the-art ones. 


\end{abstract}
\begin{keywords}
speech enhancement, magnitude reconstruction, phase estimation, collaborative learning.
\end{keywords}
\section{Introduction}
\label{sec:intro}

Signals contaminated by various background interference in daily life vastly degrade the speech quality for human listeners.  In this context, speech enhancement (SE) came into being, aiming to improve hearing comfort by separating human voice and noise signals. Traditional SE methods~\cite{boll1979suppression,hu2013cepstrum,gerkmann2011unbiased} have been extensively studied in the last decades, achieving remarkable improvements in general but breaking down when faced with non-stationary scenarios. Recently, this issue has been well addressed with the involvement of deep learning paradigms.

\begin{figure}[ht]
	\centering
	\includegraphics[width=0.48\textwidth]{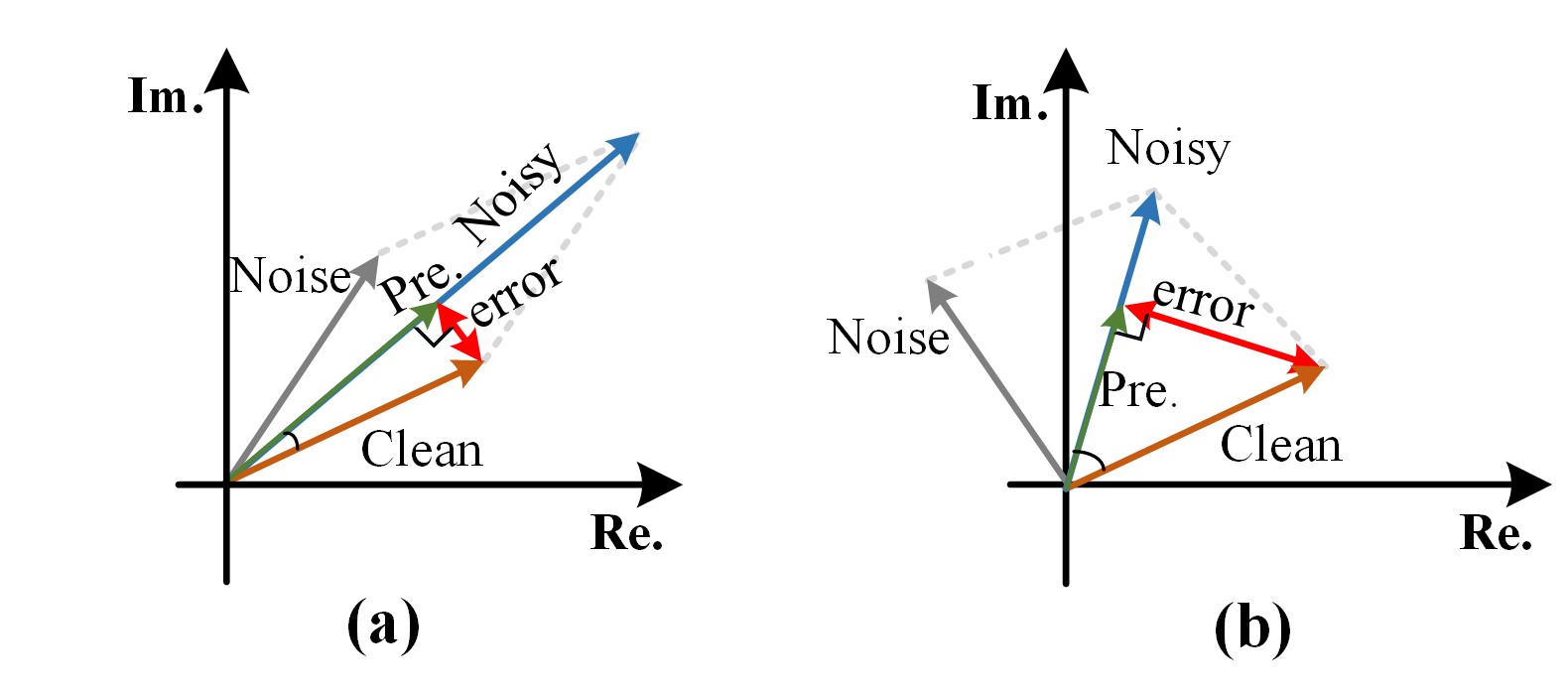}
	\label{fig1:phase}
	\caption{Phase influence under different conditions. (a) is under the surroundings with the same energy for noise and clean parts in similar directions. (b) is under the surroundings with the same energy for noise and clean parts in the opposite direction.}
\end{figure}

According to the training patterns they work on, existing single-channel SE methods can be classified into two categories: waveform-based time-domain speech estimation and time-frequency (T-F) domain spectrum reconstruction. As the spectrum contains more distinct feature patterns, suppressing noise in the T-F instead of the time domain seems to have more advantages~\cite{li2021icassp,yin2020phasen}.


Most previous studies explore the recovery of magnitude spectrogram and directly incorporate the noisy phase for speech waveform reconstruction~\cite{tan2018gated,lan2020redundant,tai2021idanet}. However, as shown in Fig.~\ref{fig1:phase}, the unprocessed phase restricts the performance upper bound of magnitude-based SE methods. Based on this observation, Yin et al.~\cite{yin2020phasen} proposed a two-stream framework, utilizing magnitude information to facilitate phase prediction. 

In this paper, we argue that phase information appears irregular, signifying that operating on phase spectrum is intractable to model phase information accurately, regardless of the assistance of external knowledge. Considering the pros and cons of each training pattern, i.e., magnitude spectrum-based methods can take advantage of apparent spectral regularity, whereas complex spectrum-based methods can implicitly estimate phase information, we present a unified two-branch framework that fosters strengths and circumvents weaknesses of different paradigms. In concrete, we split a SE task into two sub-targets, one branch for direct magnitude spectrum reconstruction and the other for implicit phase estimation. 

\begin{figure*}[ht]
	\centering
	\includegraphics[width=0.8\textwidth]{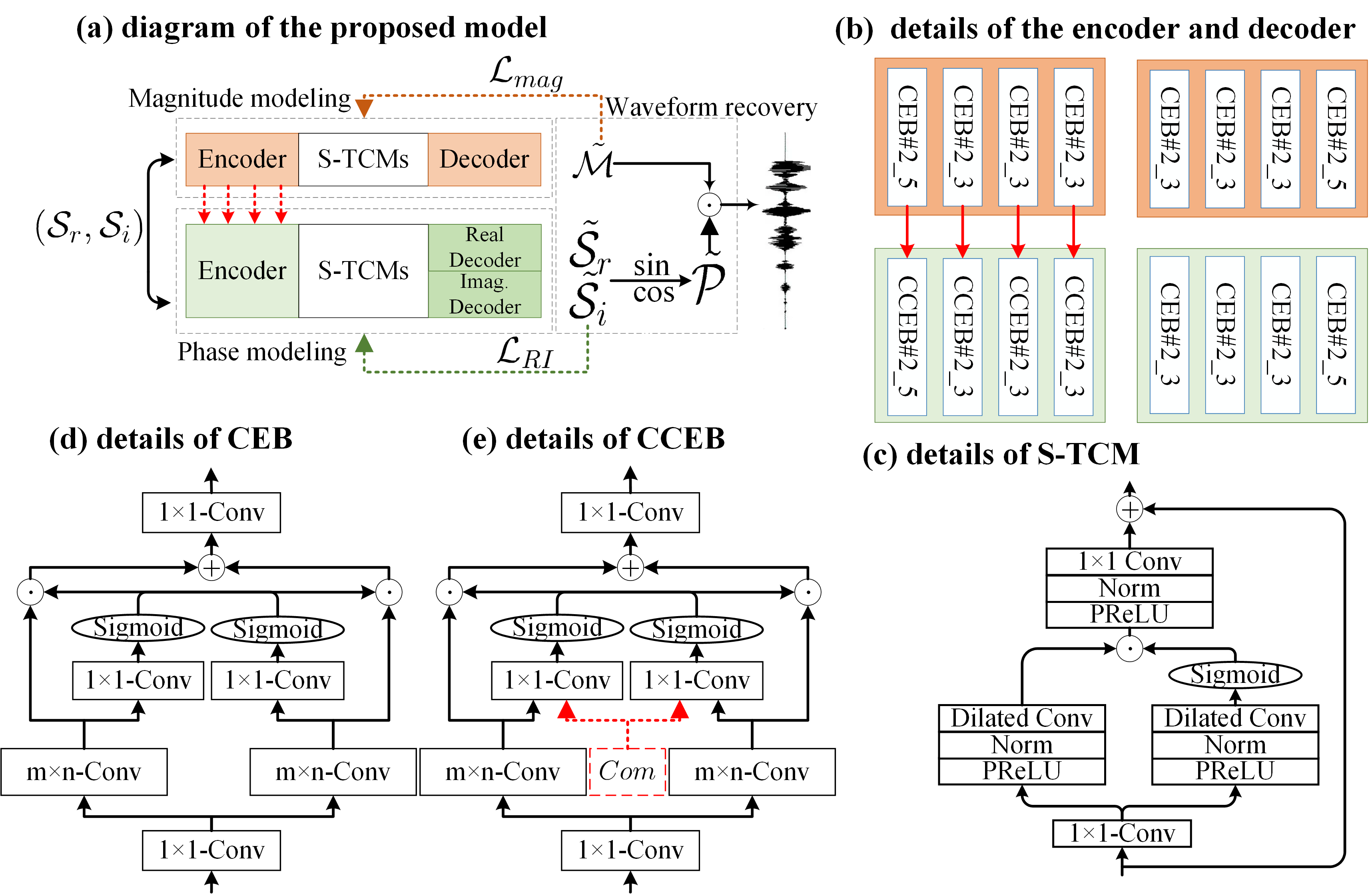}
	\caption{(a) refers to the diagram of the proposed model. (b) is the detail of encoder and decoder, where $\#m \_ n$ denotes the kernel size of the convolution. (c) is the detail of S-TCM. (d) (e) are flowcharts of the CEB and CCEB, respectively. $Com$ refers to the compensatory feature from the magnitude-based branch, and it influences the gating operation by direct addition operation. To meet the real-time requirement as well as alleviate the parameter burden, we set the stride to (1, 2) for encoder and decoder. For simplicity, we omit normalization and activation in (d) (e).}
	\label{fig:model}
\end{figure*}

In the design of different branches, we utilize the information from the magnitude-based stream as the additional supervised signal to facilitate the feature processing of the complex-based stream. Meanwhile, the implicit estimated phase from the complex-based stream will serve as a substitute for the noisy one. Besides, inspired by expert learning~\cite{jain2019multi}, we replace the regular convolution layer with our proposed collaborative expert block (CEB) and its variants to increase the model's capability of feature processing. Comprehensive experiments on TIMIT dataset verify the effectiveness and superiority of our framework.


\section{Methodology}
\label{sec:meth}

We design a speech enhancement system with two objectives in mind. First, it should make full use of the merits of each training paradigm. To realize this, we propose a two-branch SE framework to recover magnitude and phase spectrum simultaneously. Second, each branch should avoid their corresponding drawbacks as much as possible. Considering the vocals are easily distinguishable on the magnitude spectrum, and there exist potential associations between magnitude and complex spectrum, we use external knowledge from the magnitude-based stream to calibrate feature processing of the complex-based stream. In addition, the phase information implicitly optimized by the complex-based stream will unify with the estimated magnitude spectrum from the magnitude-based stream to reconstruct the speech signal.

\subsection{Notations}
We present the diagram of the proposed approach, as shown in Fig.~\ref{fig:model}. In this paper, we denote the $(\mathcal{S}_r, \mathcal{S}_i)$ as the noisy complex spectrum after short time fourier transform (STFT), while $(\tilde{\mathcal{S}}_r, \tilde{\mathcal{S}}_i)$ are the estimated real and imaginary parts from the complex-based stream. Then, estimated (complex) phase $\tilde{\mathcal{P}} \in \mathbb{R} ^{T \times F \times 2}$ can be obtained from $(\tilde{\mathcal{S}}_r, \tilde{\mathcal{S}}_i)$ via trigonometric function. Finally, estimated magnitude $\tilde{\mathcal{M}} \in \mathbb{R} ^{T \times F \times 1}$ from the magnitude-based stream is used along with $\tilde{\mathcal{P}}$ to reconstruct the time-domain speech waveform by inverse STFT. 

\subsection{The proposed two-branch framework}


Both sub-networks use complex spectrum as input and have a similar network topology, consisting of a standard encoder-decoder framework. Instead of using a regular convolutional layer as the basic unit for encoder and decoder, CEB and its variants are utilized to better capture the intrinsic harmonic features of the spectrum. In the bottleneck layer, we use stacked temporal convolution modules (S-TCM) proposed in~\cite{li2021two} to capture both short and long-range sequence dependencies. Fig.\ref{fig:model} (c) presents the architecture of S-TCM. It first squeezes the input feature into a lower dimension through a $1 \times 1$ convolution layer. A self-gating operation is applied to better control the information flow, followed by a $1 \times 1$ convolution layer. Each branch in the self-gating part uses 1-D convolution blocks with increasing dilation factors. We use 3 groups of S-TCMs as the sequential module, each of which stacks 6 S-TCM units with different dilation rates, namely, (1, 2, 4, 8, 16, 32).


\subsection{Collaborative expert block (CEB)}
With the growing demand for speech-related services over the past few years, the scenarios requiring speech enhancement are becoming more diverse and complex. This observation increases concerns about current SE systems: do we need to train a separate model for each scenario to guarantee the denoising performance? Considering that training multiple models is cumbersome from a commercialization point of view, we get inspiration from expert systems and design CEB (Fig.~\ref{fig:model} (d)) as a substitute for the regular convolutional layer. Note that for the encoder in the complex-based branch, we use a variant of CEB dubbed CCEB (compensatory and collaborative expert block) to utilize additional supervised signals from the magnitude-based stream.


We use two parallel convolutions as different experts. Different experts are expected to have different views on the same object. Each expert's output is used to offer guidance and control the other's information flow via the gating mechanism. To ease the parameter burden, we squeeze and excite features (channel dimension $64 \rightarrow 32$ and $32 \rightarrow 64$) before and after expert learning by applying $1 \times 1$ convolution layers.

\subsection{loss function}

Since two sub-networks have different training targets, we apply different loss functions to optimize sub-networks until convergence. Specifically, we train magnitude-based branches with mean absolute errors (MAE) loss in magnitude and train complex-based branches with MAE losses in real and imaginary parts. Two branches are trained jointly, and the total loss is defined as:
\begin{equation}\label{equ-1}
	\left \{
	    \begin{aligned}\mathcal{L}_{mag}&=\frac{1}{N} \sum_{i=1}^N \lvert \tilde{\mathcal{M}}-\mathcal{M}\rvert\\
		\mathcal{L}_{RI}&=\frac{1}{N} \sum_{i=1}^N (\lvert \tilde{\mathcal{S}}_r-\mathcal{S}_r \rvert + \lvert \tilde{\mathcal{S}}_i-\mathcal{S}_i \rvert)\\
		 \mathcal{L}&=\alpha \cdot \mathcal{L}_{mag} +(1-\alpha) \cdot \mathcal{L}_{RI}
	    \end{aligned}\right.
\end{equation} 
where $N$ refers to the number of training samples. $\tilde{M}$ and $M$ represent the enhanced magnitude and clean magnitude. $\tilde{S}_r$ and $\tilde{S}_i$ denote the real and imaginary part of the ground truth. $\alpha$ is the weighted coefficient, and we set $\alpha$ as 0.5 in this work.

\section{Experiments}
\label{sec:expe}
\subsection{Dataset}
We conducted our experiments using the TIMIT corpus~\cite{garofolo1993darpa}. 2000, 100 and 192 clean utterances are randomly selected for training, validating and testing. We create training and validation datasets under SNR levels ranging from -5 dB to 10 dB with the interval 1 dB, and generate the testing dataset under the SNR conditions of (-5 dB, 0 dB, 5 dB, 10 dB). 100 non-speech noises from ~\cite{hu2010tandem} and 5 life noises (cafeteria, restaurant, park, office, and meeting) from ~\cite{thiemann2013demand} are used for training and validation. Other five types of noises (babble, f16, factory2, m109, and white) from ~\cite{varga1993assessment} are used to evaluate the generalization capacity of all networks. All noises are first concatenated into a long vector. Then, an excerpt of this vector having the same length as the clear utterance is randomly chosen, scaled, and subsequently mixed with the given clean signal, providing the desired SNR condition. As a result, we generate approximately 36 hours of data for training, 1.5 hours for validation, and 1 hour for testing.

\subsection{Experimental setting}
All utterances are sampled at 16 kHz. A 20 ms Hamming window is utilized, with 50\% overlap in adjacent frames. 320 point FFT is used, leading to 161-D spectral features. The batch size is set to 2, and utterances are zero-padded to match the length of the longest in a mini-batch. All networks are trained for 30 epochs, and the learning rate is fixed at 0.0002. We optimize models by Adam. The Short-Time Objective Intelligibility (STOI) ~\cite{taal2010short} and Perceptual Evaluation of Speech Quality (PESQ)~\cite{rix2001perceptual} are selected for speech quality evaluation.

\begin{table*}[ht]
	\caption{Objective results comparisons among different models in terms of STOI and PESQ. The best value of each case is marked in bold. $p$ denotes the number (million) of parameters.}
    \label{tab:table1}
	\centering
	\begin{tabular}{c|cc|cc|cc|cc}
		\hline Test SNR & \multicolumn{2}{|c|}{$-$5dB} & \multicolumn{2}{|c|}{0dB}  & \multicolumn{2}{|c|}{5dB} &\multicolumn{2}{|c}{10dB} \\
		\hline Metric &STOI(\%)&PESQ&STOI(\%)&PESQ&STOI(\%)&PESQ&STOI(\%)&PESQ\\
		\hline Noisy &60.82&1.43&71.89&1.76&81.83&2.15&89.09&2.47\\
        \hline CCRN($p=17.44$)&73.15&1.81&82.61&2.27&88.96&2.71&93.90&3.07\\
        GCRN ($p=9.77$)&73.67&1.91&82.96&2.34&89.17&2.73&94.07&3.09\\
		PHASEN ($p=5.05$)&74.76&2.11&82.33&2.44&87.96&2.73&91.75&2.97\\
		CTS-Net ($p=4.35$)&76.16&2.04&84.65&2.45&90.69&2.87&94.77&3.25 \\
		\hline
		\bf{Ours} ($p=3.21$)&\bf78.97&\bf{2.30}&\bf{86.71}&\bf{2.69}&\bf{91.96}&\bf{3.06}&\bf{95.38}&\bf{3.37}\\
		\hline
	\end{tabular}
\end{table*}

\subsection{Baselines}
\label{sec:baselines}
We compare our proposed model with several baselines including CCRN~\cite{tan2019complex}, GCRN~\cite{tan2019gcrn}, PHASEN~\cite{yin2020phasen}, and CTS-Net~\cite{li2021two}, all of which consider phase optimization. Specifically, CCRN is a complex spectrum-based model that incorporates convolution and LSTM. GCRN is developed on CCRN by replacing regular convolutions with gated linear units. PHASEN explicitly estimates the phase with the assistance of the magnitude spectrum. CTS-Net is a two-stage approach that successively models the magnitude and complex spectrum. 

We use complex mapping MAE loss for pure complex domain methods (CCRN, GCRN), while we use complex mapping MAE loss and magnitude mapping MAE loss for the others (PHASEN, CTSNet, ours). Besides, as PHASEN is originally utilized to spectrogram under 512 point FFT, we extend it to 320 FFT for fair comparisons.

\subsection{Experimental results}
Table~\ref{tab:table1} shows the SE performance of ours and existing state-of-the-art models. We can see methods that take additional consideration of the magnitude spectrum yield better results than methods that merely use the complex spectrum under low SNR conditions. When SNR is high, the influence of magnitude tends to be small. The rationale behind this phenomenon is: as SNR degrades, the clearer structure of harmonics in magnitude spectrum can offer a better direction for network optimization. Our proposed model outperforms the previous best model PHASEN and CTS-Net on both metrics. Compared with PHASEN, we adopt a roundabout way to obtain phase from real and imaginary spectrums rather than directly estimate phase information, which avoids the difficulty brought by irregular texture. In addition, unlike CTS-Net, which simply supplies the magnitude knowledge at the beginning of the phase estimation, we provide a more fine-grained information-sharing scheme by utilizing synchronous knowledge to calibrate information at each layer.





\subsection{Visualizing phase influence}

\begin{figure}[ht]
	\includegraphics[width=0.45\textwidth]{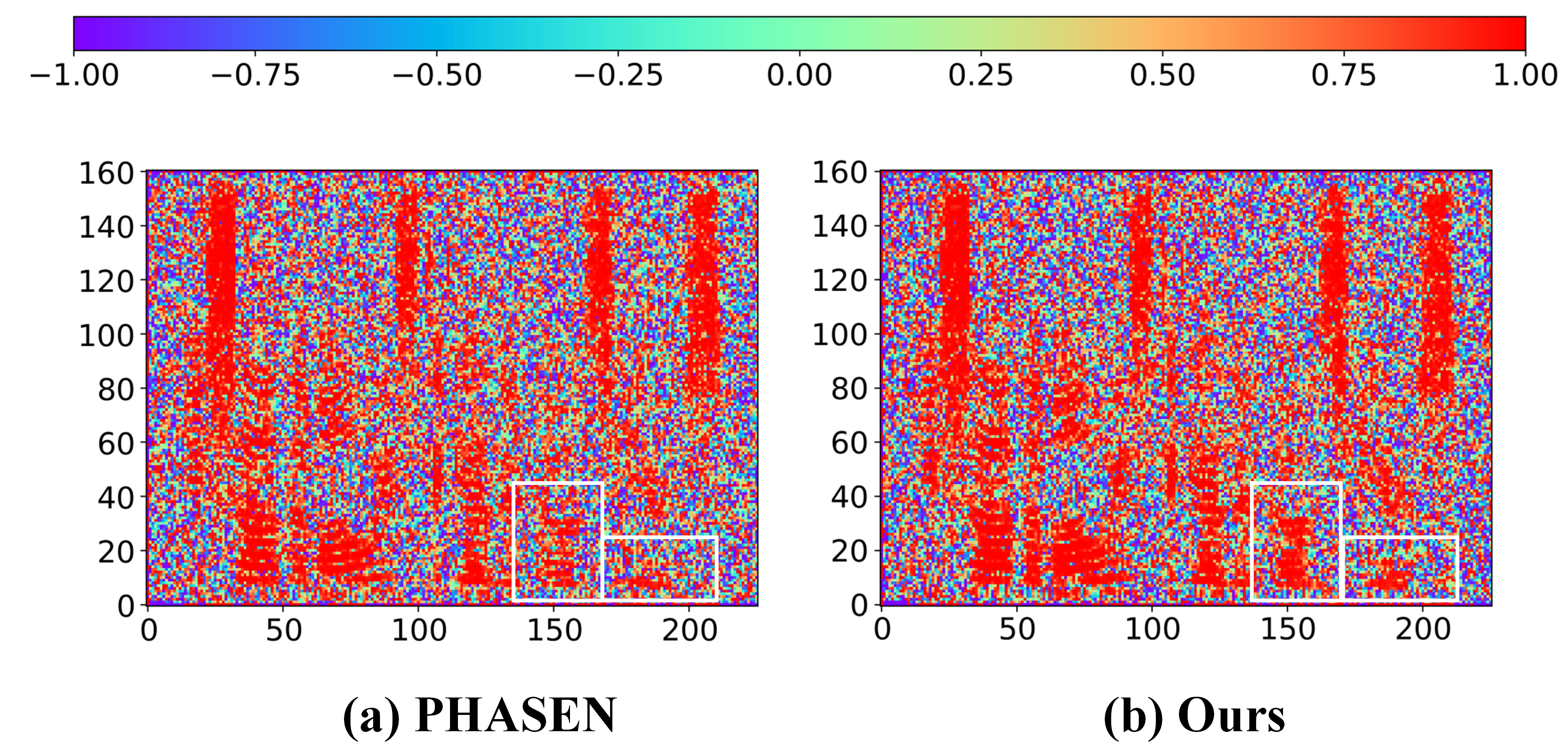}
	\caption{Visulizations of $cos(\mathcal{P}_{cl} - \mathcal{P}_{est.} )$ spectrum: (a) PHASEN, (b) ours. $\mathcal{P}$ denotes the phase spectrum.}
	\label{fig:phase}
\end{figure}

The benefit of our framework in phase prediction can be visualized in Fig.~\ref{fig:phase}, where we provide figures of phase difference to illustrate the compensation of the magnitude-based stream for explicit and implicit phase estimation. With the assistance of external knowledge from the magnitude-based stream, both PHASEN and our model have the ability to identify texture borders. The phase spectrum estimated by our framework is much closer to the ground truth, especially those areas between harmonics (white box), efficiently optimizing the human auditory perception of estimated speech. This observation verifies our assumption that exploring the potential regularity from the chaos phase spectrum is much difficult than that from the complex spectrum.


\subsection{Ablation study}
In an ablation study, we compare our framework with three variants to investigate the effects of each part. Among these variants, w/o phase represents that we use estimated magnitude and unaltered phase to restore speech; w/o multi-experts represents that we replace each CEB/CCEB with a regular convolutions; w/o compensation represents that we forgo interactions from magnitude-based streams to complex spectrum-based streams. As shown in Table.~\ref{tab:tab2}, we can observe that each part can indeed improve the SE performance.

\begin{table}[ht]
    \centering
	\caption{Objective results comparisons. $m$ refers to million.}
	\label{tab:tab2}
	\begin{tabular}{c|c|c|c}
		\hline Model&Param.($m$)&STOI(\%)&PESQ \\
		\hline w/o phase &1.60&87.24&2.78\\
        w/o multi-experts &3.41&87.51&2.71\\
        w/o compensation &3.21&87.96&2.83\\
        Ours& 3.21&88.25&2.86\\
		\hline
	\end{tabular}
\end{table}


\section{Conclusion}
\label{sec:con}
We proposed the two-branch framework with collaborative learning for monaural speech enhancement. Considering the apparent spectral regularity and potential associations between magnitude and complex spectrum, we use external knowledge from the magnitude-based branch as assistance to facilitate complex spectrum reconstruction. Meanwhile, implicit phase information from the complex spectrum-based branch will replace the noisy one. By doing so, advantages from magnitude-based methods and complex-based methods can be well incorporated.
\label{sec:method}
\bibliographystyle{IEEEbib}
\bibliography{main.bib}

\end{document}